\documentclass[aps,nofootinbib,citeautoscript,eqsecnum,notitlepage,11pt]{revtex4-1}
\pdfoutput=1

\newcommand{\bk}{{\bf k}}
\newcommand{\bq}{{\bf q}}

\newcommand{\beq}{\begin{eqnarray}}
\newcommand{\eeq}{\end{eqnarray}}
\newcommand{\beqq}{\begin{eqnarray*}}
\newcommand{\eeqq}{\end{eqnarray*}}

\newcommand{\be}{\begin{equation}}
\newcommand{\ee}{\end{equation}}

\newcommand{\bp}{{\bf p}}

\newcommand{\ve}{{\varepsilon}}
\newcommand{\nn}{\nonumber \\}

\renewcommand{\vec}[1]{{\bf #1}}

\newcommand{\sgn}{\mbox{sgn}}

\renewcommand{\epsilon}{\varepsilon}

\usepackage{amsthm, amssymb}
\usepackage{float}
\usepackage{array}
\usepackage{blindtext}

\usepackage{amsmath,amssymb,bm} 
\usepackage{graphicx}

\usepackage[tight]{subfigure} 

\usepackage{color} 
\usepackage[papersize={8.5in,11in}]{geometry}
\usepackage[colorlinks=true]{hyperref}
\hypersetup{
    bookmarks=true,         % show bookmarks bar?
    unicode=false,          % non-Latin characters 
    pdftoolbar=true,        % show Acrobat
    pdfmenubar=true,        % show Acrobat 
    pdffitwindow=false,     % window fit to page when opened
    pdfstartview={FitH},    % fits the width of the page to the window
    pdfkeywords={keyword1} {key2} {key3}, % list of keywords
    pdfnewwindow=true,      % links in new window
    colorlinks=true,       % false: boxed links; true: colored links
    linkcolor=magenta, %red,          % color of internal links (change box color with linkbordercolor)
    citecolor=blue,        % color of links to bibliography
    filecolor=magenta,      % color of file links
    urlcolor=blue           % color of external links
} 

\begin{document}

\title{Search for plasmons in isotropic Luttinger semimetals}

\author{Ipsita Mandal$^{1,2}$}

\affiliation{$^1$Laboratory ofAtomic And Solid State Physics, Cornell University, Ithaca, NY 14853,}
\affiliation{$^2$Kavli Institute for Theoretical Physics, University of California, Santa Barbara, CA 93106-4030}

\date{\today}

\begin{abstract}
Luttinger semimetals include materials like gray tin ($\alpha$-Sn) and mercury telluride, which are three-dimensional gapless semiconductors having a quadratic band crossing point (QBCP).
Due to a growing interest in QBCPs and new experimental efforts, it is essential to study the finite-temperature properties of such systems. In this paper, we investigate the emergence of plasmons in the presence of Coulomb interactions in isotropic Luttinger semimetals, for zero as well as generic nonzero temperatures. When the Fermi level lies right at the QBCP, which is the point where twofold degenerate conduction and valence bands touch each other quadratically, we find that plasmons cannot appear at zero temperature. However, for nonzero temperatures, thermal plasmons are generated. Whether they are long-lived or not depends on the values of temperature, effective electron mass and effective fine-structure constant, and the number of fermion flavors. We also numerically estimate the behavior of the inelastic scattering rate at nonzero temperatures, as a function of energy, where the signatures of the QBCP thermal plasmons show up as a sharp peak. Our results will thus serve as a guide to experimental probes on these systems.
\end{abstract}

\pacs{}

\maketitle

\tableofcontents
%==========================================================================

\section{Introduction}
A three-dimensional (3d) system with a quadratic band crossing point (QBCP)
is an example of semimetals possessing a Fermi point. They have gained widespread attention in current research~\cite{MoonXuKimBalents,Herbut,Herbut2,Herbut3,LABIrridate,rahul-sid,ipsita-rahul,ips-qbt-sc}, as pyrochlore iridates $\text{A}_ 2\text{Ir}_ 2\text{O}_ 7$ (A is a lanthanide element~\cite{pyro1,pyro2}) have been shown to host a QBCP. It has also been realised that in 3d gapless semiconductor bandstructures, in the presence of a strong-enough spin-orbit coupling, the Fermi level can coincide with a QBCP~\cite{Beneslavski}. The resulting model of a semimetal is indeed relevant for materials like gray tin ($\alpha$-Sn) and mercury telluride (HgTe). These systems are also known as ``Luttinger semimetals"~\cite{igor16}, because the low-energy fermionic degrees of freedom are captured by the Luttinger Hamiltonian  of inverted band-gap semiconductors \cite{luttinger,murakami}.
Another interesting aspect of QBCP semimetals is that the long-range nature of the
Coulomb interaction drives the ground state of such a
system to a non-Fermi liquid. This was argued by Abrikosov~\cite{Abrikosov} in 1971, and re-examined closely more recently by Moon \textit{et al}~\cite{MoonXuKimBalents}. Hence, in addition to quantum critical Dirac systems, this seems to be a simple instance of emergent non-Fermi liquid behavior, as most other well-studied cases involve the presence of a finite Fermi surface~\cite{nfl0,nfl1,nfl2,sur,ips-sc-isn,ipsc2,ips2,ips-fflo}.

The aim of the current work is to study the inelastic electron-electron scattering resulting from Coulomb interaction effects in QBCPs, at finite energies and/or nonzero temperatures ($T$). These properties are important in determining experimentally measurable quantities like conductivity, and spectral properties of clean samples at low temperatures. Furthermore, these properties can be directly probed in
transport, angle-resolved photoemission
spectroscopy (ARPES) and scanning tunneling microscopy (STM) experiments. Such experimental investigations of these systems have just begun \cite{armitage}. We believe this work will prove useful in guiding these experiments.

Abrikosov~\cite{Abrikosov} employed two different methods to obtain a controlled theory of the Coulomb-interaction mediated non-Fermi liquid behavior of the QBCP semimetal: (a) dimensional regularization, (b) expansion in $1/N_f$, where $N_f$ is the number fermion flavors (or  QBCP points at the Fermi level).
We adopt the second approach here, and assume that the Coluomb interaction can be treated within the random phase appoximation (RPA) for sufficiently large $N_f$. We must mention here that such a large $N_f$ expansion breaks down in the presence of a finite Fermi surface~\cite{lee}, with the 3d case corresponding to a marginal non-Fermi liquid~\cite{nfl1,nfl2,ips2,ips-sc-isn}.

The paper is organized as follows. In Sec.~\ref{model}, we explicitly write down the Hamiltonian for isotropic Luttinger semimetals, harboring a QBCP. In Sec.~\ref{secpol}, we compute the bare polarization bubble, at both zero temperature limit as well as at a generic temperature. In Sec.~\ref{secCol}, we treat the Coulomb interaction within RPA, and examine the emergence of plasmons. There, we also numerically compute the scattering rate from the electron self-energy. Appendix~\ref{calcu} is devoted to the description of the steps employed to evaluate the integrals involved in the $T>0$ case.

%%%%%%%%%%%%%%%%%%%%%%%%%%%%%%%%%%%%%%%%%%%%%%

\section{Model}
\label{model}

In our model for 3d isotropic quadratic band crossings, the low energy bands form a four-dimensional representation of the lattice symmetry group \cite{MoonXuKimBalents}, and can be captured by the the Luttinger Hamiltonian, with parameters corresponding to an inverted band structure and full rotational symmetry.  
This describes a spin-orbit coupled system with total angular momentum $J=3/2$, and a quadratic dispersion.
The standard $\vec{k} \cdot \vec{p}$ Hamiltonian can be written by using the five $4\times 4$ Euclidean Dirac matrices $\Gamma_a$ as \cite{Herbut}:
 \begin{equation}
 \mathcal{H}_0 = \vec d_\vec{k} \cdot \vec \Gamma \,.
\label{bare}
 \end{equation}
%%%%%%%%%%%%%%%%
Here, the $\,\Gamma_a$ forms one of the (two possible) irreducible, four-dimensional Hermitian representations of the five-component Clifford algebra, defined by the anticommutator $\{ \,\Gamma_a, \,\Gamma_b \} = 2\, \delta_{ab}$. The five components of the vector $\vec   d_{\vec{k}}$ are the real $\ell=2$ spherical harmonics, with the following structure:
 \begin{eqnarray}
\label{ddef}
  {d}^1_\vec{k} &=& \frac{\sqrt{3}\, k_y \,k_z  } {2\,m}  \,,
\quad  d^2_\vec{k} =  \frac{\sqrt{3}\, k_x\, k_z} {2\,m} \, ,\quad
 d^3_\vec{k} = \frac{ \sqrt{3} \,k_x\, k_y} {2\,m} \, ,
\nonumber\\
  d^4_\vec{k}  &=&\frac{\sqrt{3}  \,  (k_x^2 - k_y^2) }{4\,m}\,, 
\quad  d^5_\vec{k}   = \frac{2\, k_z^2 - k_x^2 - k_y^2}{4\,m} \,,
\end{eqnarray}
with $m$ being the effective electron mass and $\vec k$ denoting the 3d electron momentum vector.
The magnitude of $\vec d_{\vec{k}}$ is $d_{\vec{k}}=\frac{k^2}{2\,m}$. The energy eigenvalues of $ \mathcal{H}_0 $ are $\pm d_{\vec k}$, and hence the system has a QBCP at $k=0$.

In $d=3$, the space of $4\times 4$ Hermitian matrices is spanned by the identity matrix, the five $4\times 4$ Gamma matrices $\Gamma_a$ and the ten distinct matrices $\Gamma_{ab} = \frac{1}{2\,i}\, [\Gamma_a, \Gamma_b]$. The five anticommutating gamma-matrices can always be chosen such that three are real and two are imaginary \cite{igor12}. We choose a representation in which $(\Gamma_1, \Gamma_2, \Gamma_3)$ are real and $(\Gamma_4, \Gamma_5 ) $ are imaginary. We also note that
\begin{align}
\sum \limits_{a} \Gamma_a\,\Gamma_a = 5\,.
\end{align}

%%%%%%%%%%%%%%%%%%%%%%%%%

\section{Bare polarization bubble}
\label{secpol}

\begin{figure}
{
\includegraphics[width = 0.15 \linewidth]{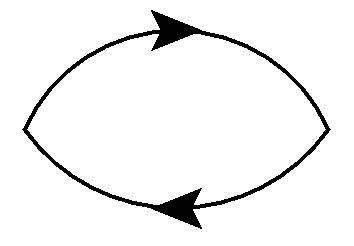}
}
\caption{\label{feyn1}Bare polarization bubble.}
\end{figure}
%%%%%%%%%%%%%%%%%%%%%%%%%%%%%%%%%%%%%%%%%%%%%%%%%%%%%%%%%%%%%%%
Screening is a many-body property directly
related to the polarizability of the electrons around the Fermi surface for a metal. In QBCP, because the density of
states (DOS) vanishes at the band touching point, the  polarization function describes the susceptibility of the
vacuum to particle-hole pair production.

In this section, we will calculate the bare polarization bubble ( \textit{i.e.}\ without any interaction line in the loop) shown in Fig.~\ref{feyn1}. This will help us determine if a plasmon mode can exist when we add Coulomb interactions to the system. After analytically continuing to real frequencies, the bubble is given by the expression~\cite{abrikosov-book}:
%%%%%%%%%%%%%%%%%%%%%%
\begin{align}
\text{Re} \, \Pi^R (\omega,\bq) &= -\frac1{(2\,\pi)^{4}}\int d^3\bp \, d\ve \, 
\tanh\left( \frac{\ve}{2\, T}\right ) \text{Tr}  \left[ G''(\ve, \bp)
\, G'(\ve - \omega,\bp - \bq) + G''(\ve, \bp - \bq)\, G'(\ve+\omega, \bp)   \right], \nonumber \\
\text{Im} \, \Pi^R ( \omega, \bq) &= \frac1{(2\pi)^{4}} \int d^3\bp \, d\ve \, \left[ \tanh\left(\frac{\ve}{2\,T}\right ) - \tanh \left(\frac{\ve - \omega}{2\,T}   \right) \right] \text{Tr} \left[ G'' (\ve, \bp) \, G'' (\ve - \omega, \bp - \bq) \right],
\label{pol1}
\end{align}
where the integration range of $p$ has the ultraviolet cut-off $\Lambda$.
The advanced and retarded fermionic Green's functions have the forms:
\begin{align}
&G^R(\ve, \bp) = -\frac12 \left\{ \frac{I + \frac{ \vec d_{\vec p} \cdot \boldsymbol{\Gamma}}
{ d_{\vec p}}}
{\ve -  d_{\vec p} + i\,0^+}  + \frac{I - \frac{ \vec d_{\vec p} \cdot \boldsymbol{\Gamma}}{ d_{\vec p}}}{\ve +   d_{\vec p} + i\,0^+} \right\}\,,
\nn
& G^A(\ve, \bp)=\lbrace G^R(\ve, \bp) \rbrace^*
= -\frac12 \left\{ \frac{I + \frac{  \vec d_{\vec p} \cdot \boldsymbol{\Gamma}}{ d_{\vec p}}}{\ve -  d_{\vec p} - i\,0^+}  
+ \frac{I - \frac{ \vec d_{\vec p} \cdot \boldsymbol{\Gamma}}{ d_{\vec p}}}{\ve +   d_{\vec p} - i\, 0^+} \right\}\,,
\end{align}
respectively,
leading to
\begin{align}
%%%%%%%%%%%%
%& G^R(\ve, \bp) = \frac {\ve + \vec d_{\bp}.\vec \Gamma + i \,\delta}
%{ \vec d_{\bp}^2 -\ve^2 }\,,\nn
%%%%%%%%%%%%%%%%
& G'(\ve, \bp) =\frac{ G^R(\ve, \bp) + G^A(\ve, \bp)}{2}
= \mathcal{P}  \left (  \frac{\vec d_{\bp}.\vec \Gamma +\ve }  {\vec d_{\bp}^2-\ve^2}\right ),\nn
& G'' (\ve, \bp)=\frac{G^R(\ve, \bp) - G^A (\ve, \bp)} {2\,i}
= \frac{\pi \left [ \left(  d_{\bp} + \vec d_{\bp}\cdot \Gamma \right) \delta (\ve-d_{\bp}) + \left(  d_{\bp} - \vec d_{\bp}\cdot \Gamma \right) \delta (\ve+d_{\bp}) \right ]}
{2\,  d_{\bp}}
\label{gexp}
\,,
\end{align} 
where the symbol $\mathcal{P}(f)$ is used to indicate that we should take the pricipal value while integrating over the function $f$. 
%%%%%%%%%%%%%%%%%%%%%%%%%
Without loss of generality, we take $\vec q $ along the $z$-axis and denote the angle between $\vec q$ and $\vec k$ by $\theta$. After taking the trace in the gamma-matrix space, we evaluate the dot products by using the relation $ \vec d_\vec p \cdot \vec d_\vec{k} =\frac{3\,(\vec{p}\cdot\vec{k})^2 - p^2\, k^2} { 8 \,m^2}
=\frac{  \left (  3\,\cos^2  \theta_{\vec p \vec k} - 1  \right ) p^2\,k^2 } { 8 \,m^2}$  ( derived in Ref.~\onlinecite{igor16}),
where $\theta_{\vec p \vec k} $ is the angle between $\vec p$ and $\vec k$.

%%%%%%%%%%%%%%%%%%%%%%%%%
\subsection{Zero temperature limit}
The zero temperature calculation is easy to perform and we can get analytical expressions for the polarization bubble.
The answers will tell us if we can get a plasmon mode at zero temperature.
To accomplish our goal, we perform the $ p \equiv |\vec p|$ integrals of Eq.~(\ref{pol1}) first. This gives us:
\begin{align}
\text{Im} \, \Pi^R ( \omega \geq 0, \bq) = &
 \frac{m^{3/2} \,T \, \sqrt{\epsilon }} {\sqrt{2} \, \pi }
\int d\epsilon \,d(\cos \theta ) \left[ \tanh \left(\frac{\epsilon }{2\,  T}\right)-\tanh \left(\frac{\epsilon -\omega }{2\, T}\right)\right ] 
 \nn &  \hspace{ 1.5  cm} \times
 \Bigg [ \left(\frac{3\, q^2 \left(\cos ^2 \theta  -1\right)}
 {-2\, \sqrt{2 \,m\, \epsilon}\, q \, \cos \theta +2 \,m \,\epsilon +q^2}+4\right)
 \delta \left(\frac{q^2}{m}-\frac{2\, \sqrt{2 \, \epsilon } \cos  \theta\, q}{\sqrt{m}}+2 \,\omega \right)
 %%%%%%%%%%%
 \nn & \hspace{ 2.2 cm } -\frac{3 \,q^2 \left(\cos ^2\theta -1\right) 
 \delta \left(\frac{q^2}{m}-\frac{2\, \sqrt{2 \,  \epsilon }\, \cos \theta\, q}{\sqrt{m}}+4\, \epsilon -2\, \omega \right)}
 {-2 \,\sqrt{2 \,m\, \epsilon}\, q 
 \, \cos \theta +2\, m\, \epsilon +q^2}  \, \Bigg ]\,,
 %%%%%%%%%%%%
\nn \text{Im} \, \Pi^R ( \omega, \bq) = & \, \text{Im} \, \Pi^R (- \omega  , \bq) \,.
\end{align}
%%%%%%%%%%%%%%%%%%%%%%%%%%%%%%%%%%%%%%%%%%%%%%%%%%%%%%%%%%%%

To take the zero temperature limit, we perform a Sommerfeld expansion in large $1/T$. The leading order term is then given by:
\begin{align}
\text{Im} \, \Pi^R ( \omega , \bq) = &
3 \,\Theta  \left( |\omega| -\frac{q^2}{4\,m}\right)\,\sgn(\omega)
\frac{  8\, m^{3/2}\, \sqrt{4 \,|\omega| -\frac{q^2}{m}}
-\frac{2 \left(q^2-2\, m \,|\omega| \right)^2 
\,\ln \left(\frac{q \sqrt{4\, |\omega| -\frac{q^2}{m}}
+2\, \sqrt{m}\, |\omega| }
{2 \,\sqrt{m}\, |\omega| -q\, \sqrt{4\,| \omega| -\frac{q^2}{m}}}\right)}{q\,| \omega| } }
{32\, \pi }\,,
\end{align}
for $T \sim 0$.
%This basically gives the result for the zero-temperature limit.
Using the variable $\tilde z = \frac{m\,z}{q^2}$, we can express the above as:
\begin{align}
\text{Im} \, \Pi^R ( z , \bq) = &
\frac{3\, m\, q \,\sgn(z)\,
f_1( \frac{m\,z}{q^2} )}
{32 \pi }\,,
%%%%%%%%%%%%
\nn  f_1(\tilde z )
= &  \, \Theta  \left( |z| -\frac{1}{4}\right)
\left[ 8 \,\sqrt{4\,| \tilde{z}|-1}
-\frac{ \left(2\, |\tilde{z}|-1\right)^2 } { |\tilde z| }
\, \ln \left(\frac{\left(2\, |\tilde{z}| +\sqrt{4\, |\tilde{z}|-1}\right)^2}
{\left(\sqrt{4 \,|\tilde{z}|-1}-2\,| \tilde{z}|\right)^2}\right) \right ].
\end{align}

We can now evaluate $\text{Re} \, \Pi^R ( \omega, \bq)$
by using Kramers-Kronig relations. Using the fact that $\text{Im} \, \Pi^R ( z < 0, \bq)
= \text{Im} \, \Pi^R ( z \geq 0, \bq)$, we get:
\begin{align}
\text{Re} \, \Pi^R ( \omega \geq 0, \bq) = &
\int_{1/4}^{\infty}\, dz\,\text{Im} \, \Pi^R ( z \geq 0, \bq)
\left( \frac{1}{z-\omega} -\frac{1}{z+ \omega} \right),
%%%%%%%%%%%%%%%%%%%%%%%
\nn \text{Re} \, \Pi^R ( \omega, \bq) = &  - \text{Re} \, \Pi^R (- \omega  , \bq) \,.
\end{align}
After performing some cumbersome calculations, we finally obtain:
\begin{align}
\text{Re} \, \Pi^R ( \omega \geq 0, \bq) = &
\begin{cases}
& \frac{ 3\,  m \,q\, f_2 \left (\frac{m\,\omega }{q^2} \right )} { 8  }    \text{ for } \omega < \frac{q^2}{4\,m} \\
%%%%%%%%%%%%%%%%%%%%%%%
&\frac{3\, m\,q\, f_3 \left (\frac{m\,\omega }{q^2} \right )} { 8 }    \text{ for } \omega \geq \frac{q^2}{4\,m}
\end{cases}\,,
\label{zero-real-pi}
\end{align}
where
\begin{align}
f_2 ( \tilde \omega  )
=&
\frac{
(1-2 \,{\tilde \omega})^2 \,\cot ^{-1}\left(\frac{2\, {\tilde \omega}}{\sqrt{1-4\, {\tilde \omega}}}\right)
-2 \,{\tilde \omega} \left(\sqrt{4\, {\tilde \omega}+1}+\sqrt{1-4\, {\tilde \omega}}-4\, \pi \right)
-2\, (2 \,{\tilde \omega}+1)^2 \tan ^{-1}\left(\sqrt{4\, {\tilde \omega}+1}\right)
}
{\tilde{\omega }} \,,
\nn
f_3( \tilde \omega  )
= & \frac{  (2\, \tilde \omega+1)^2 \left[\pi -2\, \tan ^{-1}\left(\sqrt{4 \,{\tilde \omega}+1}\right)\right]
-2\, {\tilde \omega}\, \sqrt{4\, {\tilde \omega}+1} }
{ \tilde{\omega }} 
\,.
\end{align} 
Ref.~\onlinecite{Beneslavski} found that the polarization bubble (1) evaluates to zero for zero momentum, and (2) is of the form $\tilde m \,q$ for $|\omega| \ll q$ (where $\tilde m$ is of the order of $m$).
Our results are thus consistent with their studies, since $\text{Re} \, \Pi^R ( \omega \geq 0, \bq)
\Big  \vert_{\omega \ll q}
\simeq \frac{3}{2} \left (\pi -2\right ) m\, q $.

The final expressions tell us that $\text{Re} \, \Pi^R ( \omega \geq 0, \bq)> 0$ for all $\omega $, and hence we do not expect to find a zero temperature plasmon. This was to be expected because the Fermi surface in this case is just a point.
The process of creation of particle-hole pairs involves incoherent excitations of electrons from the lower to the
upper band. However, there is no phase space for intraband excitations at zero temperature due to the Pauli principle.

\subsection{Generic temperature}
%%%%%%%%%%%%%%%%%%%%%%
Since we did not find a plasmon mode at $T=0$, let us check if it can appear at a finite temperature, which is anyway the realistic scenario in experiments. For finite $T$, we expect to see a crossover in the behavior of the
polarization function, because a nonzero $T$ will effectively act as a nonzero chemical potential. This results in a nonvanishing DOS, making intraband excitations possible. These excitaions dominate the infrared
behavior of the polarization function and their collective modes can give rise to long-lived plasmons. 

The finite temperature calculation of the polarization bubble is very non-trivial and a full analytical expression is not possible to derive. We will instead try various tricks to simply the integrals and evaluate them in some discrete regimes.
To evaluate the integrals of Eq.~(\ref{pol1}) at a generic temperature, we use the elliptic coordinates defined as:
\begin{align}
& p=\frac{ q \left ( \xi +\eta \right ) } {2}\,,\quad 
|\vec p-\vec q| = \frac{ q \left ( \xi +\eta \right ) } {2}\,,\quad
\cos \theta = \frac {1 + \eta\, \xi}  {\eta + \xi }\,,
%%%%%%%%%%%%%%%%%%
\end{align}
\footnote{ This implies that
$  p_x =\frac{ q \,\sqrt{(1-\eta^2)\,(\xi^2 -1 ) }} {2} \cos \phi
\,,\quad p_y =\frac{ q \,\sqrt{(1-\eta^2)\,(\xi^2 -1 ) }} {2} \sin \phi\,,
\quad p_z = \frac{ q \,\left ( 1+\eta\,\xi \right ) } {2} 
%%%%%%%%%%%%%%%%%%%%%%
$, and hence the integral measure is: $
   \int   d^3\vec p 
 \rightarrow   \int  \frac{q^3\,\left( \xi^2-\eta^2   \right) } {8 }\,  d\xi \,d\eta\,d\phi
 \quad \left [\text{ the Jacobian is } \frac{q^3\,\left( \xi^2-\eta^2   \right) } {8}
 \right ].
$
}
with $1 \leq \xi \leq \frac{ 2\,\Lambda }{q}$, $-1\leq \eta \leq 1 $ and $0 \leq \phi \leq 2\,\pi$.
This gives us:
\begin{align}
&\text{Re} \, \Pi^R (\omega,\bq) =   
   t_1^R+t_2^R  \,, \nn
%%%%%%%%%%%%%%%%%%%%%%%%%%%
\nn
&t_1^R= \frac{ 2\, m\, Q \sqrt{ m\, T}} {\pi^2}
\int_{-1}^1 d\eta\, \int_1^{\frac{ 2\,\Lambda }{q}} d\xi\,
\frac{ \eta\, \xi \left [3 + \eta^4 - 
   3 \,\xi^2 + \xi^4 + \eta^2 \left ( \xi^2-3  \right ) \right ] 
 \left [\tanh[ \frac {Q^2 \left(\eta - \xi\right )^2} {2}] - 
   \tanh \left ( \frac { Q^2 \left(\eta + \xi \right )^2} {2}\right )   \right ]
   }
   {  \left (\xi^2-\eta^2 \right ) \left ({\tilde \beta}^2 - 4\, \eta^2 \,\xi^2\right )}\,,
\label{pol1a}\\
%%%%%%%%%%%%%%%%%%%%%%%%%%%%%%%%%%%%%%%5
\nn
&t_2^R= -\frac{ 3\, m\, Q \sqrt{ m\, T}} {\pi^2}
\int_{-1}^1 d\eta\, \int_1^{\frac{ 2\,\Lambda }{q}} d\xi\,
\frac{ \left(  1-\eta^2 \right) \left(  \xi^2 -1 \right) 
\left(  \eta^2 +\xi^2 \right) 
 \left [\tanh[ \frac {Q^2 \left(\eta - \xi\right )^2} {2}] +
   \tanh \left ( \frac { Q^2 \left(\eta + \xi \right )^2} {2}\right )   \right ]
   }
   {  \left (\xi^2-\eta^2 \right ) \left ({\tilde \beta}^2 - 4\, \eta^2 \,\xi^2\right )}\,,
\label{pol1b}
%%%%%%%%%%%%%%%%%%%%%%555
\end{align}
and
\begin{align}
& \text{Im} \, \Pi^R ( \omega, \bq)=   t_1^I+t_2^I  \,, \nn
%%%%%%%%%%%%%%%%%%%%%%%%%%%
\nn
&t_1^I= \frac{  m\, Q \sqrt{m\, T}  } {\pi}
 \int_1^{\frac{ 2\,\Lambda }{q}} d\xi\,
\frac{ \left [ {\tilde \beta}^4 + 4 \, {\tilde \beta}^2 \xi^2 \,( \xi^2-3) + 
      16 \,\xi^4 \left (3 - 3 \,\xi^2 + \xi^4\right )\right ]
 \sinh \left(\frac{Q^2 \,{\tilde \beta}}  {2}  \right )  
  }
  {  \xi^3 \left ( 4 \,\xi^4-{\tilde \beta}^2   \right ) 
  \left  [\cosh \left(\frac{Q^2 \,{\tilde \beta}}  {2} \right ) + 
   \cosh \left (\frac{ Q^2 \,{\tilde \beta}^2 } {8 \,\xi^2} +\frac{Q^2\,\xi^2}{2}\right )  \right ]
   }\,   \Theta\left ( 4 \,\xi^2 - {\tilde \beta}^2 \right ) \,,
\label{pol2a}\\
&t_2^I= - \frac{ 12\, m\, Q \sqrt{m\, T} } {\pi}
 \int_0^{1} d\eta\,
\frac{  \left (1 - |{\tilde \beta}|+ \eta^2  \right ) \left (1 - \eta^2 \right )
 \sinh \left(\frac{Q^2 \,{\tilde \beta}}  {2}  \right )  
  }
  {   
\left (|{\tilde \beta}| - 2\,\eta^2 \right ) \sqrt{ |{\tilde \beta}| - \eta^2}  
  \left  [
  \cosh \left(    \frac{Q^2 \,{\tilde \beta}}  {2} \right ) + 
  \cosh \left  ( Q^2\,\eta \,\sqrt{   |{\tilde \beta} | -\eta^2}  \right )
 \right ] 
   }\,   \Theta \left ( {\tilde \beta}^2 - 1 - \eta^2 \right ) \,,
 \label{pol2b}
\end{align}
where we have written the expressions in terms of the dimensionless variables
$\vec Q =\frac{\vec q}{ 2 \,\sqrt{m\, T}}$, $ \Omega = \frac{\omega }{T}$, $ {\tilde \beta} = \frac{\Omega }{Q^2}$. 
%We note that $  \Pi^R (\Omega, 0 ) = 0 $.

\begin{figure}
\subfigure[] {
\includegraphics[width = 0.4 \linewidth]{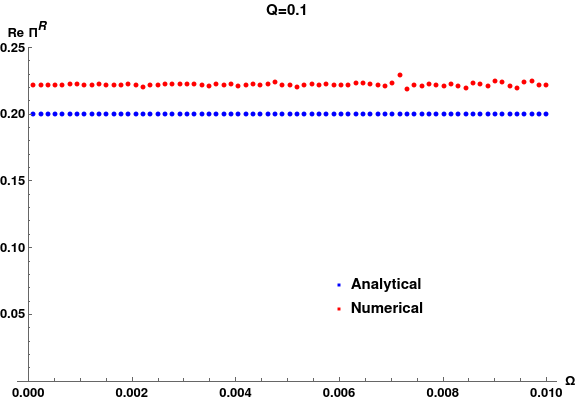}
}  \quad
\subfigure[] {
\includegraphics[width = 0.4 \linewidth]{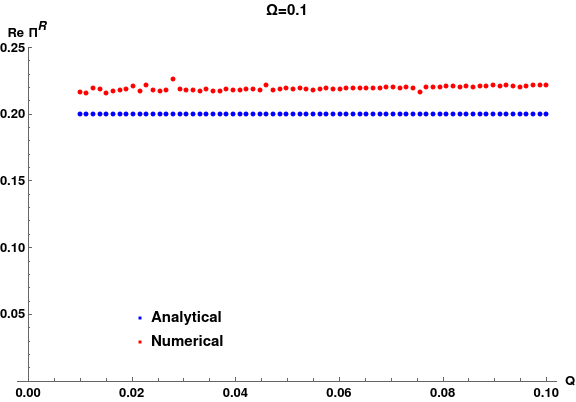}
}  
\subfigure[] {
\includegraphics[width = 0.4 \linewidth]{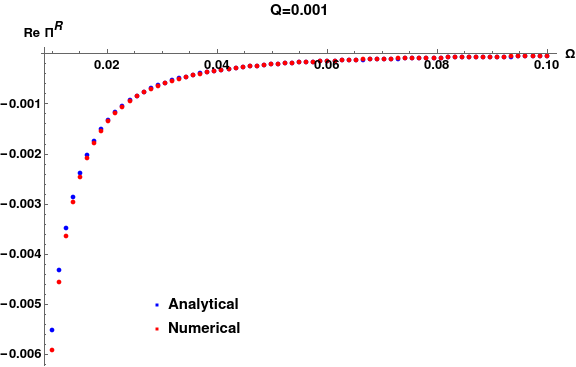}
}  \quad
\subfigure[] {
\includegraphics[width = 0.4 \linewidth]{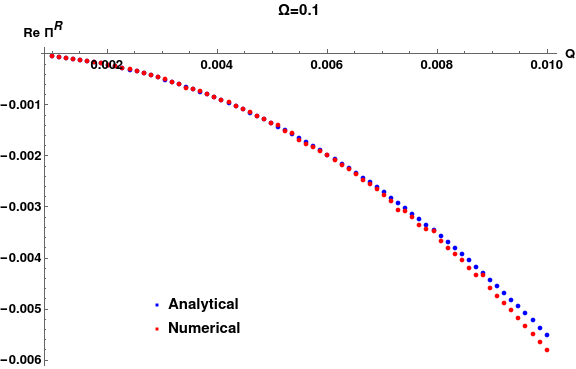}
}  
\subfigure[] {
\includegraphics[width = 0.4 \linewidth]{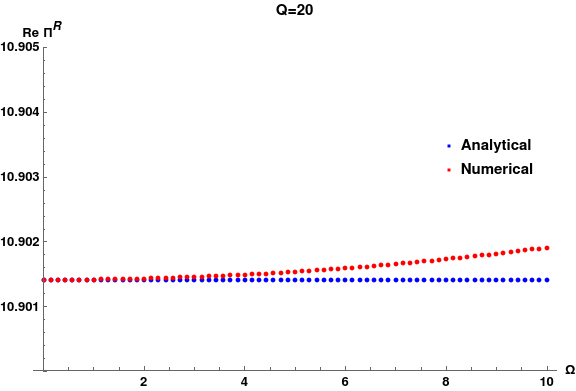}
}  \quad
\subfigure[] {
\includegraphics[width = 0.4 \linewidth]{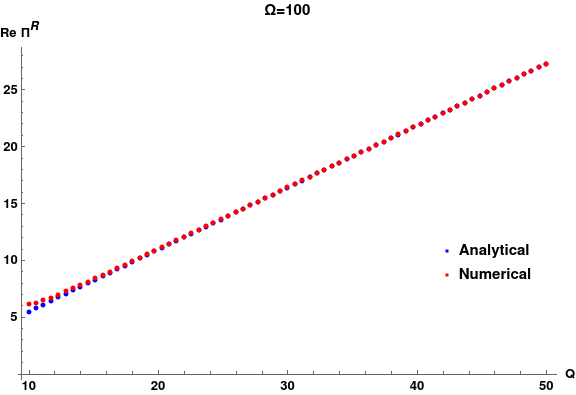}
}  
\caption{\label{figrepi}In these plots, we have compared our analytical approximations of $\text{Re} \, \Pi^R (\omega, q)$ to the numerical results for various regimes. We have set $m=T=1$.}
\end{figure}

\begin{figure}
\subfigure[] {
\includegraphics[width = 0.4 \linewidth]{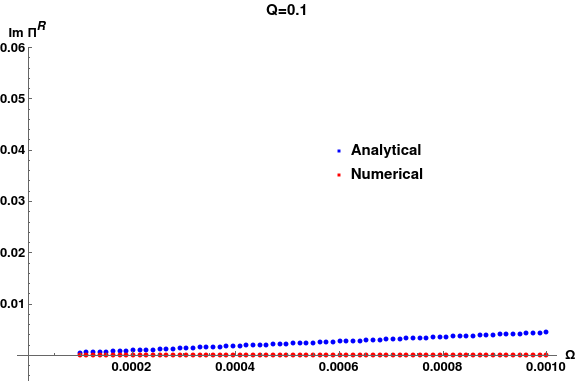}
}  \quad
\subfigure[] {
\includegraphics[width = 0.4 \linewidth]{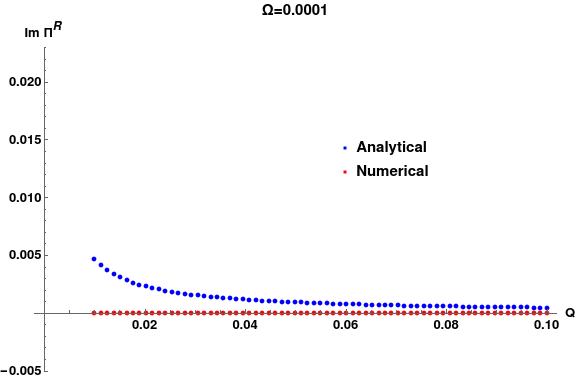}
}  
\subfigure[] {
\includegraphics[width = 0.4 \linewidth]{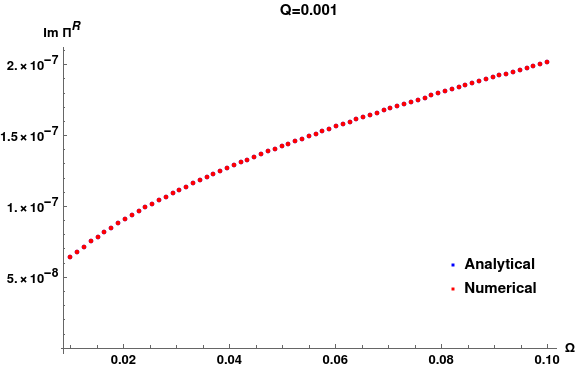}
}  \quad
\subfigure[] {
\includegraphics[width = 0.4 \linewidth]{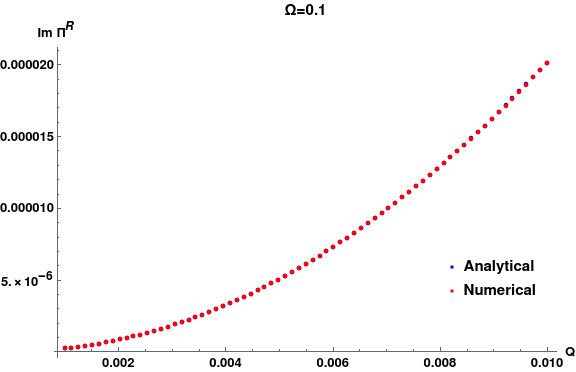}
}  
\subfigure[] {
\includegraphics[width = 0.4 \linewidth]{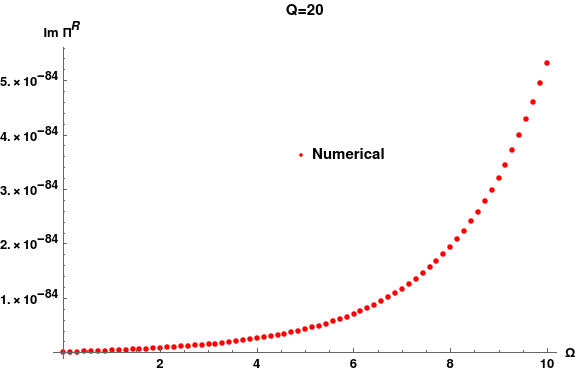}
}  \quad
\subfigure[] {
\includegraphics[width = 0.4 \linewidth]{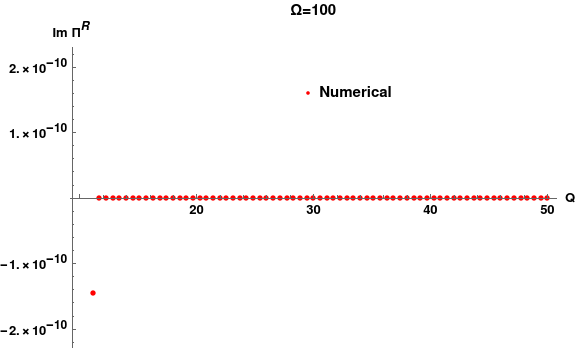}
}  
\caption{\label{figimpi}In these plots, we have compared our analytical approximations of $\text{Im} \, \Pi^R (\omega, q)$ to the numerical results for various regimes. We have set $m=T=1$. In (c) and (d), analytical and numerical points completely overlap on one another. In the last two plots ((e) and (f)), the value of $\text{Im} \, \Pi^R (\omega, q)$ is zero for all practical purposes, and in this region we have used zero in our analytical expressions (not shown on these two graphs).}
\end{figure}

Performing the integrals analytically is very difficult. We state the approximate results in some limiting cases:
\begin{align}
&\text{Re} \, \Pi^R (\omega, q) =
 \begin{cases} 
%%%%%%%%%%%%%%%%%
0.2\, m \, \sqrt{m\,T}
 & \mbox{ for } Q\ll 1 \text{ and }   |\Omega | \ll Q^2 \,, \\
%%%%%%%%%%%%%%%%%%%%%%%%%%%%%%%%%%%%%%%%%%%
-\frac{ Q^2 \,m\, \sqrt{m\,T}
\left( 0.389348\, Q^2+0.547416  \,  \Omega ^2\right)}
{\Omega ^4 }
 = & -\frac{q^2\, T^{3/2} \left(0.136854 \,m+
\frac{0.0243343\, q^2\, T}  { \omega ^2}  \right)}{\sqrt{m}\, \omega ^2} 
\mbox{ for } Q\ll 1 \text{ and } |\Omega | \ll 1
\\ & \hspace{ 5.3 cm} \text{ and } |\Omega | > Q \, , \\
%%%%%%%%%%%%
-\frac{Q^2 m\, \, \sqrt{m\, T}\, \left [  270  \left( \pi \, \Omega ^{3/2}  + 4 \, Q \,\Omega \right ) - 401 \right] }
{ 135 \,\pi ^2\, \Omega ^2}  
  = & \frac{q^2 \left[ \sqrt{m} \left(401\, T^{3/2}-270 \,\pi \, \omega ^{3/2}\right)-540\, q\, \omega \right] }
{540\, \pi ^2\, \omega ^2}
   \mbox{ for } Q\ll 1 \text{ and } |\Omega | \gg 1 \,, \\  
%%%%%%%%%%%%%%%%%
 \frac{ Q \, m\,  \sqrt{m\, T} \,( 3 \,\pi -6 )}
  { 2 \,\pi} = \frac{m\, q \,( 3 \,\pi -6 )}
  { 4 \,\pi} 
    & \mbox{ for } Q\gg1 \text{ and } 
 |\Omega| \ll  Q^2  \, (\text{ which implies } {\tilde \beta} \ll 1)\,,\\  
%%%%%%%%%%%%%%%%%
  \frac{16\,Q^2 \,  m\,  \sqrt{m\, T} }
  { 3 \,\pi^2 \,\sqrt {|\Omega|} }= \frac{4\, \sqrt m\, q^2   }
  { 3 \,\pi^2 \,\sqrt {  |\omega|} }
     & \mbox{ for } Q\gg1 \text{ and } |\Omega| \gg Q^2 \, (\text{ which implies } {\tilde \beta} \gg 1)
  \,,\\
\end{cases}        
\label{Pi1}
\end{align}
%%%%%%%%%%%%%%%
and
\begin{align}
&\text{Im} \, \Pi^R (\omega, q) =  \begin{cases}
%%%%%%%%%%%%% 
\frac{ \Omega\,m\,  \sqrt{m\, T} \left[  Q^2\left (
0.95493\, \ln Q -0.151526 \right )+0.463015\right ]}{Q}
= & \frac{0.92603\, m\, \omega  \left[  q^2 \left(0.515604
\, \ln \left(\frac{q}{2\, \sqrt{m \,T}}\right)-0.0818151\right)+m\, T\right]}
{q\, T}\\
  & \mbox{ for } Q\ll 1 \text{ and } |\Omega | \ll Q^2\,, \\
%  %%%%%%%%%%%%%%%%%%%%%%%%%%%%%%%%%%%%%%%%%%
\frac{ 8  \,Q^2 \,m \,\sqrt{m\,T} \, \tanh (\Omega/4 )  }
  {  \pi \,\sqrt{ |\Omega|} }
  = \frac{2\,\sqrt m \,q^2 \tanh \left(\frac{\omega }{4\,T}\right)}{\pi\,  \sqrt{ |\omega| }}  
& \mbox{ for } Q\ll 1 \text{ and } |\Omega | \ll 1 \text{ and } |\Omega | > Q \,, \\
%%%%%%%%%%%%%%%%%
\frac{ 8  \,Q^2\, \sgn (\Omega) \,m \,\sqrt{m\,T}  }
  {  \pi \,\sqrt{ |\Omega|} }
  = \frac{ 2 \, \sqrt{ m} \,q^2\, \sgn (\omega)   }
  {  \pi\,\sqrt{  |\omega|} }  
  & \mbox{ for } Q\ll 1 \text{ and } |\Omega | \gg 1 \,, \\
 %%%%%%%%%%%%
 0
    & \mbox{ for } Q\gg1 \text{ and } 
 |\Omega| \ll Q^2  \, (\text{ which implies } {\tilde \beta} \ll 1)\,,\\  
%%%%%%%%%%%%%%%%%%%%%%%%%%% 
0
    & \mbox{ for } Q\gg1 \text{ and } 
 |\Omega| \gg  Q^2  \, (\text{ which implies } {\tilde \beta} \gg 1)\,.\\  
%%%%%%%%%%%%%%%%% 
\end{cases}
  \label{Pi2}
\end{align}
Appendix~\ref{calcu} outlines the strategy employed to obtain the above expressions. In Figs.~\ref{figrepi} and \ref{figimpi}, we have plotted $\Pi^R (\omega, q)$ in some of these regions, and compared our analytically expressions with the actual integrals computed numerically.

%%%%%%%%%%%%%%%%%%%%%%%%%%%%%%%%%%%%%%%%%%%%%%%%%%%%%%%%%%%%

%%%%%%%%%%%%%%%%%%%%%%%%%%%%%%%%%%%%%%%%%%%%%%%%%%%%%%%%%%%%%%%

%%%%%%%%%%%%%%%%%%%%%%%%%%%%%%%%%%%%%%%%%%%%%%%%%%%%%%%%%%%%

%%%%%%%%%%%%%%%%%%%%%%%%%%%%%%%%%%%%%%%%%%%%%%%%%%%%%%%%%%%%%%%

Since $\text{Re} \, \Pi^R (\omega, q) <0$ in the range where $ Q\ll 1 \text{ and } 
 |\Omega| \gg Q^2 $, we might expect to find a thermal plasmon when these conditions are satisfied. For $|\Omega|>1 $, we find that $\text{Im} \, \Pi^R (\omega, q)$ is of the same order, and does not allow the pole to appear. However, for $|\Omega| <1$, the imaginary part is smaller/subleading. This will be explained in more details in the following section. We also note that:
 \begin{align}
 \Pi^R \left  ( 0, \frac{  q}{ 2 \,\sqrt{m\, T}} \ll 1  \right ) \simeq 0.2\,m\,\sqrt{m\,T}\,.
 \end{align}

%%%%%%%%%%%%%%%%%%%%%%%%%%%%%%%%%%%%%
\section{Coulomb Interaction}
\label{secCol}

\begin{figure}
\subfigure[]{
\includegraphics[width = 0.67 \linewidth]{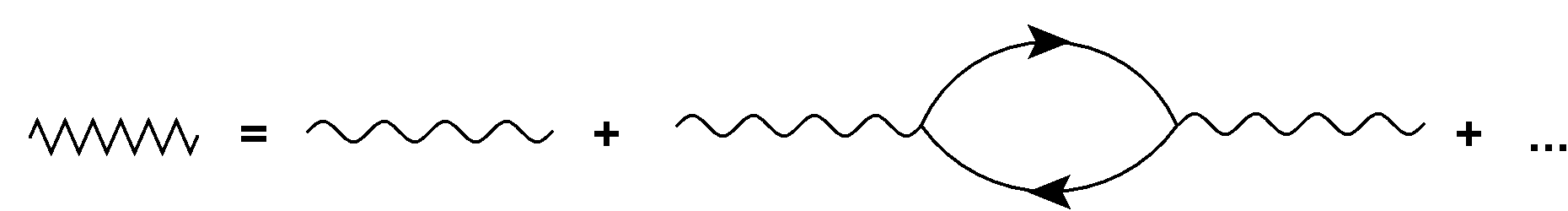}
}\\
\subfigure[]{
\includegraphics[width = 0.3 \linewidth]{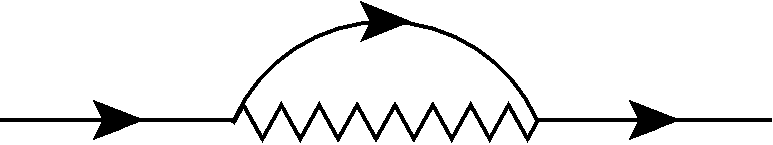}
} \hspace{2 cm}
\subfigure[]{
\includegraphics[width = 0.2 \linewidth]{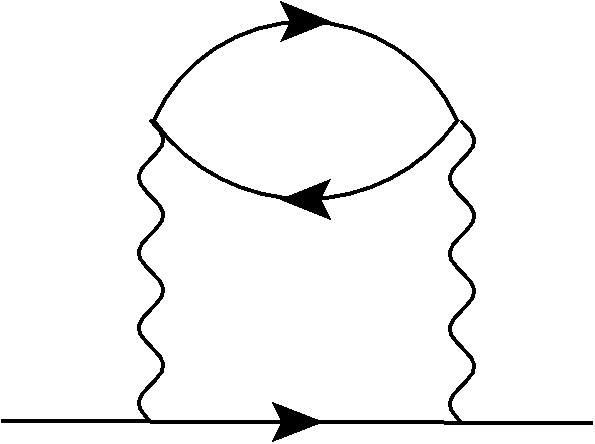}
}
\caption{\label{feyn22}The propagator for the effective Coulomb interaction $V(\omega, \bq)$, denoted by the zigzag line. The wavy line corresponds to the propagator of the bare Coulomb interaction, $V_0(\bq)$. (b) One-loop fermion self-energy. (c) The two-loop perturbative contribution to the fermion self-energy. This non-RPA second-order diagram can be neglected within the $1/N_f$ expansion.}
\end{figure}
%%%%%%%%%%%%%%%%%%%%%%%%%%%%%%%%%%%%%%%%%%%%%%%%%%%%%%%%%%%%%%%

Having obtained the results for polarization bubble in various regimes, we are now in a position to calculate the inelastic scattering rate due to Coulomb interactions.
The nonzero scattering rate should result from the imaginary part of interactions. Since the bare Coulomb interaction is real, we therefore need to take into account screening
effects, for example, within the RPA. We assume that the effective interaction  $V^R(\omega, q)$ is given by the RPA series:
%%%%%%%%%%%%%%%%%%
\be
V^R(\omega, q) = \frac{V_0(q)}{1+V_0(q)\, N_f\,  \Pi^R(\omega, q)}\,, 
\label{eqRPA}
\ee
as shown in Fig.~\ref{feyn22}(a). This is true in the large $N_f$ limit~\cite{Abrikosov}.
Here, $ \Pi^R (\omega, q)$ is the polarization bubble at one-loop (computed earlier), and $   V_0(q)= \frac{\alpha}{q^2}
\equiv \frac{4\,\pi\,e^2}{\epsilon\,q^2}$ is the bare Coulomb interaction in a material with a dielectric constant $\epsilon$ ($ e$ is the electron charge). The quantity $\frac{e^2}{\epsilon}$ is usually referred to as the effective fine structure constant. Plasmons emerge at frequencies $\omega$ where there are poles in the effective Coulomb potential (or zeros of the dielectric function $ \mathcal{E}(\omega, q) \equiv \epsilon \left[ 1+V_0(q)\, N_f\,  \Pi^R(\omega, q)\right ]$).

In the static limit of $\omega=0$ and low momenta, we have:
%%%%%%%%%%%%%%%%%%
\begin{align}
 V^R(0, \bq) = & \frac{V_0(\bq)}{1+V_0(\bq)\, N_f\,  \Pi^R(0, Q\ll 1)}
%\nn &
\simeq \frac{ \alpha}
{q^2 +0.2 \, m\, \sqrt{m\, T} \,  \alpha \, N_f}\,, 
\label{Eq:V}
\end{align}
such that the Thomas-Fermi wave-vector is given by
$
 q_{\texttt{TF}} ^2 \simeq  0.2\,  \sqrt{m^3\, T} \,  \alpha \, N_f$, which sets the size of the screening
cloud. This implies that the thermally induced screening length is given by:
 \begin{align}
\ell_{scr} \equiv \frac{1}{q_{\texttt{TF}}} \simeq \frac{2.2 }{\sqrt{\sqrt{m^3\, T} \,  \alpha \, N_f} }\,.
\end{align}

\subsection{Plasmon pole}

%%%%%%%%%%%%%%%%%%%%%%%%%%%%%%%%%%%%%
We can think of the effective interaction $V^R(\omega ,\bq)$ as the photon propagator in the medium, such that its pole, if any, gives us the dispersion of the collective photon-electron excitations, which are the plasmons. Eq.~(\ref{zero-real-pi}) shows that the zero-temperature limit does not allow for the existence of any plasmon. However, for the generic temperature case, from Eq.~(\ref{Pi1}), we find that $\text{Re} \, \Pi^R (\Omega, Q)$ is negative in the regions satisfying $Q\ll 1 \text{ and } |\Omega | \gg 1$, creating the possibility for the effective photon propagator to have a pole. The dispersion of this possible plasmon at low momenta is given by the solution of $\left [ 1+ V_0( q )\, N_f\,  \Pi^R(\omega,  q) \big \vert_{ Q\ll 1 } \right ]= 0$. Since the dielectric function is complex-valued, it follows that for a real wavevector, the roots themselves are at complex frequencies, which just tells us that the collective excitations will have a finite decay due to Landau damping. The real part of the root is proportional to the energy of the
plasmon, and the imaginary part gives its decay rate.

For the case of $|\Omega|\gg1 $, we find the solution: 
\begin{align}
\sqrt{\omega} &=\frac{\left(\frac{1}{2}-2 \,i\right)  \sqrt{m}\, \alpha  \,N_f}{\pi }
\, ,
\end{align}
for $\omega  \geq 0$. This gives the result $\omega= -\frac{\left(\frac{15}{4}+2 \,i\right) m\, \alpha^2\,  N_f^2}{\pi ^2}$, which is clearly inadmissible. 
However,  for the case of $|\Omega|\ll 1 $, we find the solution: 
\begin{align}
\omega & = \omega_{pl}+  
\frac{0.240327\, q^2 \,T ^{1/4} } {m^{5/4} \, \sqrt{  \alpha \, N_f}}
-i\,\gamma\, ,\quad 
\omega_{pl} = 0.369938 \left (m \,T^3 \right)^{1/4} \,\sqrt{\alpha \,  N_f} \,,\nn
%%%%%%%%%
\gamma & =  \frac{0.21511 \,{m}^{1/4}\,\sqrt{\alpha \,  N_f}\, \omega_{pl}^{5/2} }{T^{7/4}}
= 0.0179054 \,m^{7/8}\, {T}^{1/8} \left ( \alpha \,  N_f \right )^{7/4} \,,
\label{decay}
\end{align}
where $\omega_{pl}$ is the plasma frequency and $\gamma $ is the decay/damping rate.
Hence we conclude that there exists a plasmon pole in this nonzero temperature case. For the plasmon to be long-lived, we need $\gamma\ll \omega_{pl}$, which is possible if $ \alpha ^2\,N_f^2\, m\, <127.126 \,T$.

%%%%%%%%%%%%%%%%%%%%
%%%%%%%%%%%%%%%%%%%%%%%%%%%%%%%%%%%%%%%%%%%%%%%%%%%%%%%%%%%%

\subsection{Inelastic scattering rate}

For computing the scattering rate due to inelastic electron-electron collisions, we need to first calculate the imaginary part of the electron self-energy due to Coulomb interaction. At the one-loop order, it is given by the diagram shown in Fig.~\ref{feyn22}(b), and its analytical expression reads as~\cite{abrikosov-book}:
%%%%%%%%%%%%%%%%%5
\be
\text{Im} \, \Sigma^R(\ve, \bk) = - \frac{1}{(2\pi)^{4}}\int d^3 \bq \, \int d\omega\,  G''( \ve - \omega, \bk -\bq) \, V'' (\omega,  \bq) 
\left[  \coth\left (  \frac{\omega}{2\,T} \right ) + 
\tanh \left ( \frac{\ve - \omega}{2\, T}\right ) \right], \label{SMImSigma1}
\ee
where $V'' = \frac{V^R - V^A} {2\,i}.$ We assume that the effective interaction  $V(\omega , q)$ is given by the RPA series, as discussed in Eq.~(\ref{eqRPA}).
%%%%%%%%%%%%%%%%%%

We note that the self-energy is a matrix in the space of the $\Gamma$-matrices, and can be parametrized as $\Sigma^R(\ve,\bk) = \Sigma^R_{s}\, I + \Sigma^R_v\, \vec d_{\vec k} \cdot \Gamma $, where we have denoted the part multiplying the identity matrix as $\Sigma^R_{s}$ and the rest as $\Sigma^R_{v}$.
Since the scattering rate involves only $\Sigma^R_{s}$, we can simplify our calculations by replacing $G''(\ve,\bp)$ with
$ \frac{\pi \left [  \delta (\ve-d_{\bp}) +  \delta (\ve+d_{\bp}) \right ]}
{2}$. This leads to:
\begin{align}
& \text{Im} \, \Sigma^R_{s}(\ve,\bk) =\frac14 \sum_{j = \pm} \int \frac{d^3 \bq}{(2\, \pi)^3} \text{Im}\, V^R(\omega_j,  q)
\left [  \coth\left (  \frac{\omega}{2\,T} \right ) + 
\tanh \left ( \frac{\ve - \omega}{2\, T}\right ) \right ]
%%%%%%%%%%%%%%%%%%%%%%%%%%%%%%
\nn & = 
-\frac{1}{16 \,\pi^2}\sum_{j = \pm } \int_{-1}^1 dt \int_0^{\Lambda} 
\frac{    dq   \, q^2 \,N_f \, V_0^2(\bq) 
\, \text{Im}\, \Pi^R(\omega_i, q)
\left [  \coth\left (  \frac{\omega_j}{2\,T} \right ) + 
\tanh \left ( \frac{\ve - \omega_j}{2\, T}\right ) \right ] 
}
{\left[1 
+ N_f \, V_0( q) \,\text{Re}\, \Pi^R(\omega_j, q)   \right]^2 
+ \left[ N_f \,V_0( q)\, \text{Im}\, \Pi^R(\omega_j, q) \right]^2} 
\,, \label{SMImSigma2}
\end{align}
with $\omega_j \equiv \epsilon +j\, \frac{(\bk - \bq)^2} {2\,m} 
= \ve + j\, \frac {k^2 + q^2 - 2\, k\, q\, t}  {2\,m}$ ($j=\pm$), $t= \cos \theta_{\bk \bq},$ and $\theta_{\bk \bq}$ denoting the angle between vectors $\bk$ and $\bq$. In terms of the dimensionless variables $\vec Q =\frac{\vec q}{ 2 \,\sqrt{m\, T}},$ $ \vec y =\frac{\vec k}
{ 2 \,\sqrt{m\, T}},$ $ \Omega_j = \frac{\omega_j }{T}
= x+ 2\, j \left ( y^2 + Q^2 - 2\, y\, Q\, t \right ) ,$ $ x =\frac{ \ve}{T}$, the self-energy takes the form:
%%%%%%%%%%%%%%%%%%%%
\begin{align}
& \text{Im} \, \Sigma^R_{s}(x\, T, 2 \,\sqrt{m\, T}\,\vec y) 
 =  
-\frac{(m\,T)^{3/2}}  {128 \,\pi^2}
\sum_{ j= \pm} \int_0^{\tilde \Lambda  =  \frac{\Lambda }{ 2 \,\sqrt{m\, T}}} 
\frac{    dQ   \, Q^2 \,N_f \, V_0^2( q ) 
\, \text{Im}\, \Pi^R(\omega_i, q)
\left [  \coth\left (  \frac{\Omega_j}{2 } \right ) + 
\tanh \left ( \frac{x - \Omega_j}{2 }\right ) \right ] 
}
{\left[1 + N_f \, V_0(  q) \,\text{Re}\, \Pi^R(\omega_j, q)   \right]^2 
+ \left[ N_f \,V_0(  q)\, \text{Im}\, \Pi^R(\omega_j, q ) \right]^2}\, . \label{SMImSigma3}
\end{align}

%%%%%%%%%%%%%%%%%%%%%%%%%%%%%%%%%%%%%%%%%%%%%%%%%%%%%%%%%%%%

\begin{figure}
\subfigure[] {
\includegraphics[width = 0.44 \linewidth]{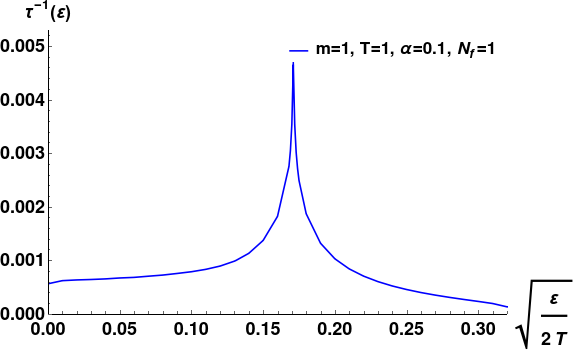}
}  \quad
\subfigure[] {
\includegraphics[width = 0.44 \linewidth]{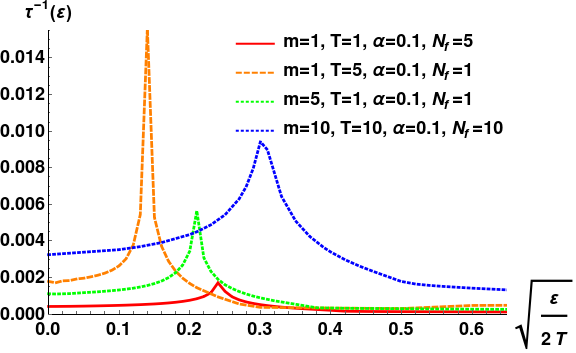}
}  
\caption{\label{figtau}Inelastic scattering rates for the QBCP show sharp peaks, indicating the existence of plasmons.
The frequency (energy) $\epsilon$ at the location of the peak gives the the plasmon frequency $\omega_{pl}$. The decay rate of the plasmon excitation is proportional to the width of the peak, which means that a sharp peak gives a long-lived plasmon. 
(a) For $\left(m=1, \,T=1,\,\alpha=0.1,\, N_f =1\right)$, the plasmon peak is seen around $\sqrt{\frac{\epsilon}{2\,T}} \simeq 0.17$.
(b) Plasmon peaks seen for some other different values of the parameters.}
\end{figure}
%%%%%%%%%%%%%%%%%%%%%%%%%%%%%%%%%%%%%%%%%%%%%%%%%%%%%%%%%%%%%%%

%%%%%%%%%%%%%%%%%%%%%%%%%%%%%%%%%%%%%%%%%%%%%%%%%%%%%%%%%%%%

\begin{figure}
  {
\includegraphics[width = 0.47 \linewidth]{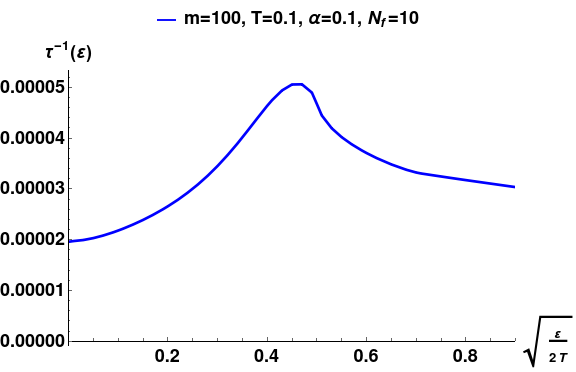}
}  
\caption{\label{figtau5}Inelastic scattering rate of the QBCP for $\left(m=100, \,T=0.1,\,\alpha=0.1,\, N_f =10\right)$ shows a shallow bump, indicating that the emergent plasmon peak is very wide, and hence short-lived.}
\end{figure}
%%%%%%%%%%%%%%%%%%%%%%%%%%%%%%%%%%%%%%%%%%%%%%%%%%%%%%%%%%%%%%%

The inelastic scattering rate, defined in the
spirit of the conventional Fermi-liquid (FL), is given by:
\begin{align}
\frac{1}{\tau (\epsilon)}
= -2\,\text{Im} \, \Sigma^R_{s}(\epsilon ,\sqrt{|\epsilon|/2}\,\hat{\vec k})\,,
\end{align}
where it implies that we are computing $\text{Im} \, \Sigma^R_{s}(x\, T, 2 \,\sqrt{m\, T}\,\vec y)$ on mass-shell, or in other words, at $|x|=2\,y^2$. Then, we have the simplified expressions: $ \Omega_+ = \frac{\omega_+ }{T}= 4\,y^2 + 2\,Q \left (  Q - 2\, y\,  t \right ) $ and  $ \Omega_- = \frac{\omega_- }{T}= 2 \,Q \left (   2\, y\,  t - Q  \right ) $. The particle-hole symmetry of the system ensures that $ \tau(\epsilon) = \tau(-\epsilon)$, and therefore, we henceforth consider only the case of $\epsilon \geq 0$.

Let us  discuss the form of the integrand for the on-shell case. Since $\text{Im}\, \Pi^R(\Omega,\vec Q)  \big \vert _ {Q\gg1 } \simeq 0 $, the fourth and fifth regions of Eq.~(\ref{Pi1}) and  Eq.~(\ref{Pi2}) give zero contributions to the integrand.
We find that $| \Omega_+ |\geq Q^2$ is always true. Hence, for $j=+$, we only need to use the expressions of  Eq.~(\ref{Pi1}) and  Eq.~(\ref{Pi2}) for the second and third regions.
Using these approximate expressions, we perform the integral in Eq.~(\ref{SMImSigma3}) numerically to analyze the variation of $\tau (\epsilon)$ as a function of the energy $\epsilon$. For the parameter values of  $(m=1, \, T=1,\,\alpha=0.1, \,N_f =1 )$, Fig.~\ref{figtau}(a) illustrates the behavior of $\tau (\epsilon)$. We notice a pronounced peak around $ y = \sqrt{\frac{\omega_{pl}}{ 4\, T}}  \simeq 0.17$, as expected, since for this value of $y$, $\omega_+ \simeq \omega_{pl}$ (in the integration region of  $Q \ll 1$ and $Q \ll |\Omega_+| $). Similar sharp peak was also found in the computation of inelastic scattering rate of 3d Dirac/Weyl semimetals~\cite{hofman,kozi-fu}, which have linear band crossing points. Fig.~\ref{figtau}(b) shows the behavior of $\tau(\epsilon)$ as a function of $
\sqrt{\frac{\epsilon}{2\,T}} $ for various values of the parameters. We have used $N_f=1$ in several cases, because the thermally induced screening of the interaction for $T>0$, is expected to restore the validity of the RPA for small enough $q$ even for $N_f \sim 1$.
%We have used $\alpha =0.1$ as the RPA for Coulomb interaction is safely applicable only for weak enough interactions. 
Lastly, Fig.~\ref{figtau5} shows the scattering rate for
$\left ( m=100, \,T=0.1,\,\alpha=0.1,\, N_f =10 \right )$, for which the plasmon is not long-lived, as can be understood from the discussion below Eq.~(\ref{decay}).

%We see another small bump, before $\tau$  finally falls off to zero, which is due to the second region contributing larger %values when $\Omega_- $ comes close to $\omega_{pl}/T$ in the integrand.

%%%%%%%%%%%%%%%%%%%%%%%
\section{Analysis and  discussion}
\label{discussion}

We have analyzed the effect of Coulomb interactions on the 3d QBCP within the RPA, valid for large $N_f$. We have found that for the case of Fermi level lying at the QBCP, although plasmons do not  exist at $T=0$, they can emerge at a finite temperature. Their decay rate depends on the values of $T$, effective electron mass and effective fine-structure constant, and the number of fermion flavors. We also note that the dispersion of the plasmons is quadratic in momentum, similar to the behavior expected in 3d Dirac/Weyl semimetals ~\cite{kozi-fu}. The non-existence of plasmons at $T=0$ is due to the vanishing of density of states at the quadratic band touching point. However, as we go to nonzero temperatures, electron-hole pairs can be excited due to thermal effects, creating the possibility of the emergence of thermal plasmons. 

The QBCP inelastic scattering rate, as a function of energy, shows a sharp peak due to the existence of the thermal plasmons. Hence, the signature of the QBCP thermal plasmons can be probed in experiments measuring transport or spectral properties. In future works, one can study the case of disordered QBCP semimetals at finite temperatures, and also the scenarios when the fully isotropic Luttinger semimetal is reduced to the ones with cubic or lower symmetry. A further generalization to be explored will be the case of unequal electron and hole masses~\cite{ipsita-rahul}, such that the conduction and valence bands have different curvatures.

\section{Acknowledgments}
We thank Michael J. Lawler and Vladyslav Kozii for helpful discussions. We are especially grateful to Erich Mueller and Matthias Punk for providing valuable insights and inputs.
We also acknowledge the warm hospitality of KITP, where this work was completed.
This research was supported in part by the National Science Foundation under Grant No. NSF PHY-1748958.

%%%%%%%%%%%%%%%%%%%%%%%%%%%%

\appendix

%%%%%%%%%%%%%%%%%%%%%%%%%%
\section{Steps for calculating the integrals at nonzero temperatures}
\label{calcu}

The integrals in Eq.~(\ref{pol1a})--(\ref{pol2b}) are quite involved and it is a tedious job to calculate the corresponding analytical solutions. Nevertheless, we obtained the approximate expressions in certain ranges of the variables $\vec Q$ and $\Omega$. 

Let us first describe the strategy employed for computing $ \text{Re} \, \Pi^R (\omega,\bq) $.
In the $Q\equiv |\vec Q| \ll 1$ limits, we divided the integration range over $\xi $ into two regions: $(1, 1/Q)$ and $(1/Q, \frac{2 \,\Lambda}{q^2} ) $, assuming $ \frac{2 \,\Lambda}{q^2} > 1/Q$. In the region  $ \xi \in (1, 1/Q)$, we expanded the expressions involving hyperbolic tangents in small $Q^2$. In the region  $ \xi \in ( 1/Q,  \frac{2 \,\Lambda}{q^2})$, we used the asymptotic expansion of the hyperbolic tangents:
\begin{align}
\label{tanh}
\tanh \left( x/2\right ) = \left ( 1-e^{-x} \right )
\left ( 1- e^{-x}  +e^{-2\,x}  - e^{-3\,x} +\ldots  \right ), \text{ for } x\rightarrow \infty\,.
\end{align}

In the $Q \gg 1$ limits, we only need to use the asymptotic expression of Eq.~(\ref{tanh}) for the terms involving hyperbolic tangents. This gave zero answer for Eq.~(\ref{pol1a}) to leading order. The dominant contribution was obtained from Eq.~(\ref{pol1b}).

Computation of $ \text{Im} \, \Pi^R (\omega,\bq)$ was significantly less complicated, as it involved only one integration (either over $\xi$ or $\eta $). For $|\Omega | \ll  1 $, we expanded $\sinh \left (  |\Omega |/2 \right )$ and $\cosh \left (  |\Omega |/2 \right )$ in small $|\Omega|$. For $|\Omega | \gg  1 $, we expanded $\sinh \left (  |\Omega |/2 \right )$ and $\cosh \left (  |\Omega |/2 \right )$ as $\frac{ e^{|\Omega |/2} }{2}$ to leading order. The expressions $\cosh \left (\frac{ Q^2 \,{\tilde \beta}^2 } {8 \,\xi^2} +\frac{Q^2\,\xi^2}{2}\right )$ and $ \cosh \left  ( Q^2\,\eta \,\sqrt{   |{\tilde \beta} | -\eta^2}  \right )$ were also approximated in a similar way, depending on the values of $Q$ and $|\Omega|$.

%%%%%%%%%%%%%%%%%%%%%%%%%%%%%%%%%%% %%%%%

\bibliographystyle{apsrev4-1}
\bibliography{biblio}
\end{document}